\documentstyle[aps,prl,preprint,floats,epsfig]{revtex}

\begin{document}

\preprint{\tighten\vbox{\hbox{\hfil CLNS 98/1542}
			\hbox{\hfil CLEO 98-3}
}}

\title{\LARGE Continuum Charged $D^*$ Spin Alignment\\ at $\sqrt{s}=10.5$ GeV} 

\author{(CLEO Collaboration)}
\date{\today}

\maketitle
\tighten

\begin{abstract}
A measurement of the spin alignment of charged $D^*$ mesons produced
in continuum $e^+ e^- \to c \bar{c}$ events at $\sqrt{s}=10.5$ GeV is 
presented. This study using 4.72 fb$^{-1}$ of CLEO II data shows that 
there is little evidence of any $D^*$ spin alignment.
\end{abstract}
\newpage

{\renewcommand{\thefootnote}{\fnsymbol{footnote}}

\begin{center}
G.~Brandenburg,$^{1}$ R.~A.~Briere,$^{1}$ A.~Ershov,$^{1}$
Y.~S.~Gao,$^{1}$ D.~Y.-J.~Kim,$^{1}$ R.~Wilson,$^{1}$
H.~Yamamoto,$^{1}$
T.~E.~Browder,$^{2}$ Y.~Li,$^{2}$ J.~L.~Rodriguez,$^{2}$
% ...
\end{center}
 
\small
\begin{center}
$^{1}${Harvard University, Cambridge, Massachusetts 02138}\\
$^{2}${University of Hawaii at Manoa, Honolulu, Hawaii 96822}\\
% ...
\end{center}

\setcounter{footnote}{0}
}
\newpage

%\pacs{13.65.+i,13.60.Le,13.87.Fh}
% 13.65.+i - Hadron production by e-e+ collisions
% 13.60.Le - Meson production
% 13.87.Fh - Fragmentation into Hadrons

\section{Introduction}

There have been numerous theoretical 
\cite{frag1,firstfrag,theory,ffth1,ffth2,ffth3,ffth4} 
and experimental \cite{charm,ffex2,ffex3,ffex4,ffex5,ffex6,ffex7,ffex8} 
studies of the fragmentation of heavy quarks. The energy 
distribution and flavor dependence of heavy quark hadronization have been 
modeled by fragmentation functions. The role that spin plays in the
hadronization process is still being investigated and is not well
understood at this time\cite{matrix,align1,align2,align3,align4,align5,zfrag}.
To increase the understanding of this role, a precise measurement of the 
probabilities of a meson being directly produced in each of the available 
spin states is needed.

At CLEO, the fragmentation of charm quarks can be analyzed by making 
measurements of primary hadrons containing charm quarks from
continuum $e^+ e^-$ annihilations. CLEO has previously published results 
of charmed meson energy distributions \cite{charm} as well as the spin 
alignment of charged $D^*$ mesons\cite{align2}. In this paper an updated 
measurement of the charged $D^*$ spin alignment using the entire CLEO II 
dataset is presented.

\section{Polarization, Alignment, and \boldmath{$P_V$}}

According to the quark model, a meson is composed of two spin 
$\frac{1}{2}$ valence quarks that can combine to form four spin states 
in the absence of orbital angular momentum, i.e. four $S$-wave states. Writing 
these in the basis of total angular momentum, $J$, and its $z$-component, 
$J_z$, they are the vector states $|1,1>$, $|1,0>$, $|1,-1>$, and the 
pseudoscalar state $|0,0>$, where the $z$-direction can be arbitrarily chosen.
The probability of an $S$-wave meson being produced in a vector state
is often described by the ratio $P_V$ defined as
$$P_V = \frac{V}{V+P} \eqno(1)$$
where $P$ and $V$ are the respective probabilities of the meson being 
created in the pseudoscalar and vector states. 

The helicity formalism is useful in the context of describing the angular 
distributions and correlations in the production and decay of particles with 
non-zero spin. For a particle with momentum $\vec{p}$, the helicity is 
defined as 
$$\lambda = \frac{\vec{J} \cdot \vec{p}}{|{\vec{p}}|}, \eqno(2)$$
which in the case of a spin-1 particle is just the $z$-component of the spin
when the $z$-direction has been chosen as the flight direction of the meson.
The helicity density matrix is often used to organize information about 
the spin of a particle.
The diagonal elements of this matrix $\rho_{\lambda \lambda}$, with
$\sum_{\lambda} \rho_{\lambda \lambda} =1$, 
represent the probability that the particle has helicity $\lambda$.

Simple statistical expectations are that all helicity states of a spin $J$ 
particle are equally populated, but production and fragmentation dynamics 
can lead
to either polarized or aligned particles. A system of particles is polarized 
if there is a net angular momentum, i.e. 
$\rho_{\lambda\lambda} \neq \rho_{-\lambda-\lambda}$ for some helicity 
$\lambda$, and it is aligned if there is a nonuniform population of states, but
$\rho_{\lambda\lambda} = \rho_{-\lambda-\lambda}$ for all $\lambda$.
Since the production and fragmentation processes in this analysis conserve 
parity and the CESR beams are unpolarized, it is expected that the $D^*$ mesons
from $e^+e^- \to \gamma^* \to c \bar{c}$ are unpolarized, but it is possible
for the $D^*$ mesons to be aligned.

To measure the spin alignment of a vector meson, the angular distribution 
of its decay products is analyzed, but because the angular distributions of 
the $\lambda =1$ and $\lambda =-1$ states are degenerate, the values of 
$\rho_{11}$ and $\rho_{-1-1}$ cannot be distinguished and only one variable, 
e.g. $\rho_{00}=1-\rho_{11}-\rho_{-1-1}$, is accessible. 
From the definition above, the vector meson is aligned if 
$\rho_{00}$ differs from $1/3$. For the case of a vector meson decaying
to two pseudoscalar mesons, the angular distribution can be written 
$$W(\cos \theta)=\frac{3}{4}[(1-\rho_{00})+(3 \rho_{00}-1)\cos^2 \theta]
\eqno(3)$$
where $\theta$ is defined as the angle of a daughter pseudoscalar in the 
parent vector meson rest frame, with respect to the direction of motion of 
the parent vector meson in the rest frame of the production process.
In our case, the production rest frame of a $D^*$ directly produced in 
charm fragmentation from $e^+ e^-$ annihilation coincides with the
lab frame.  
 
By using the variable
$$\alpha = \frac{3\rho_{00}-1}{1-\rho_{00}}, \eqno(4)$$ 
the angular distribution can be expressed as
$$W(\cos \theta)=N(1+ \alpha \cos^2 \theta)\eqno(5)$$ 
where $N$ is a normalization factor equal to $3/(6+2\alpha)$.
The value of $\alpha$ can range between $-1$ and $+\infty$, where 
the angular distribution would be isotropic if $\alpha=0$,  proportional 
to $\sin^2 \theta$ if $\alpha=-1$ and proportional to $\cos^2 \theta$ if 
$\alpha = \infty$.

Whereas the naive statistical expectation is that all 
four $S$-wave meson states are created in equal proportions, i.e. 
$\rho_{\lambda \lambda}=\frac{1}{3}$ ($\alpha=0$) and $P_V=0.75$, 
there are other 
models that have been presented where the alignment and $P_V$ vary as a 
function of momentum\cite{oldmodel,newmodel}. Heavy quark symmetry predicts
that vector mesons containing a single heavy quark are
produced unaligned, but there have been suggestions 
that the value of $P_V$ may depend upon the mass difference of the vector and 
pseudoscalar mesons\cite{theory,massmodel}.
It has also been suggested that $P_V$ is directly related to the spin alignment
\cite{link} and in
the previous CLEO $D^*$ spin alignment analysis, a value for $P_V$ was 
calculated using this relationship \cite{align2}. However, the validity of the 
statistical model is assumed when deriving this relationship. We
feel that a determination of $P_V$ for $D^*$ mesons warrants an independent
measurement which is the topic of a current CLEO analysis. The results
of the $P_V$ analysis will be presented in a future paper. 

\section{Detector and Event Selection}

The CLEO II detector is a general purpose charged and neutral 
particle detector and is described in detail elsewhere \cite{cleo}. The dataset
used in this analysis consists of 3.11 fb$^{-1}$ of data collected
at the $\Upsilon (4S)$ resonance and 1.61 fb$^{-1}$ of data collected about 
60 MeV below the resonance. This corresponds to approximately 5 million
continuum $c \bar{c}$ events.

The $D^{*+}$ in this analysis is required to decay through the 
channel $D^{*+} \to D^0 \pi^+$ with the $D^0$ decaying either through the mode 
$D^0 \to K^-\pi^+$ or $D^0 \to K^-\pi^+ \pi^0$ (inclusion of charge conjugate 
modes is implied throughout this paper).
The $\pi^+$ in the $D^{*+}$ decay is kinematically limited to having a 
momentum less than 456 MeV/$c$ in the lab frame of reference, 
and is referred to as the ``slow'' pion.

All tracks used in this analysis are required to have an impact parameter
within 5 mm of the interaction point in the plane transverse to the beam pipe
and within 50 mm in
the direction of the beam pipe. Tracks are also required to have a 
momentum less than 6 GeV/$c$ and an r.m.s. residual less than 1 mm for
their hits. Particle ID is not used 
since there is little gain for this particular analysis
and it introduces the possibility of additional systematic errors.
For a pair of photons to be considered as a candidate $\pi^0$, they must
each have an energy of at least 100 MeV, be within the ``good'' barrel
of the detector ($|\cos \theta_{detector}|<0.71$), have a shower shape
in the crystal calorimeters  consistent with that of a photon,
combine to be within 20 MeV/$c^2$ of the neutral pion mass, and have
$|\cos\theta_\gamma|<0.9$, where $\theta_\gamma$ is the decay angle of the 
photon in the $\pi^0$ rest frame, with respect to 
the $\pi^0$ direction of motion in the lab frame.

For the $D^0 \to K^- \pi^+$ mode, the $D^0$ is reconstructed by taking all 
possible pairs of oppositely charged tracks in an event, assigning the 
kaon mass to one and the pion mass to the other (or vice versa), adding their 
four-momenta, and then calculating the invariant mass. The $D^{*+}$ is 
reconstructed by adding the four-momentum of a candidate slow $\pi^+$ in the 
event to the four-momentum of the candidate $D^0$. The mass difference, 
$\Delta M$, between the candidate $D^0$ and 
$D^{*+}$ is required to be within 2.5 MeV/$c^2$ of the world-average mass 
difference of 145.42 MeV/$c^2$ \cite{PDG}.

The $D^0$ is spinless and the decay products have an isotropic angular 
distribution. However, due to the jet-like nature of continuum events, 
the background from random track combinations tends to have 
$\cos \phi_K \simeq -1$, where $\phi_K$ is the decay angle of the $K^-$ in the 
$D^0$ rest frame, relative to the $D^0$ motion in the lab frame. A 
requirement that $\cos \phi_K \geq -0.9$ is 
added to improve the signal-to-background ratio.  

For the $D^0 \to K^- \pi^+ \pi^0$ mode, the four-momentum of a candidate 
$\pi^0$ is added to the four-momenta of two oppositely charged tracks to 
form candidate $D^0$'s in the event. Mass 
difference and kaon decay angle requirements are the same as described above.

\section{Fitting}
To test models that predict that the alignment varies as a function of the 
momentum of the $D^{*+}$, the data are broken up into six $x^+$ bins in 
the range 0.25 to 1.0, where $x^+$ is a Lorentz-invariant variable defined as
$$x^+ \equiv \frac{P(D^*)+E(D^*)}{P_{max}(D^*)+E_{max}(D^*)},\eqno(6)$$
where $E_{max}=E_{beam}$, $P_{max}=\sqrt{E_{beam}^2-M_{D^{*+}}^2}$ and
$M_{D^{*+}}$ is the world-average value for the mass of a $D^{*+}$.

\begin{figure}
\centerline{\epsfig{file=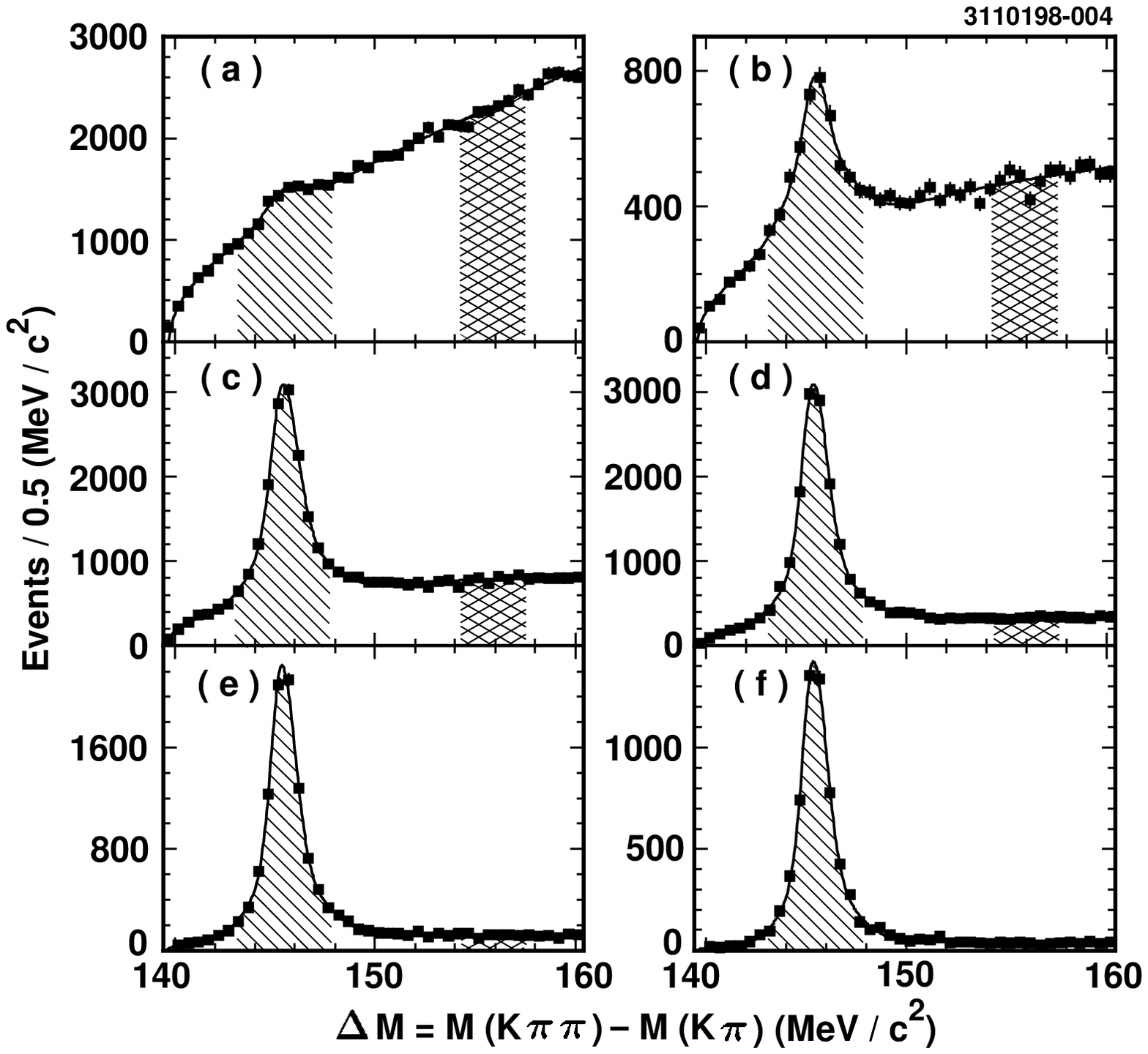,height=6.0in}}
\caption{{\label{fig:kpmsdf}}
$D^* - D$ mass difference for the $D^0 \to K \pi$ decay mode for the six 
$x^+$ ranges a) $0.25<x^+<0.45$, 
b) $0.45<x^+<0.55$, c) $0.55<x^+<0.65$, d) $0.65<x^+<0.75$, e) $0.75<x^+<0.85$,
f) $0.85<x^+<1.0$. The solid squares are the data points and the solid
line is the fitting function as described in Section IV. The hatched area
is the signal region while the cross-hatched region is the sideband.}
\end{figure}

\begin{figure}
\centerline{\epsfig{file=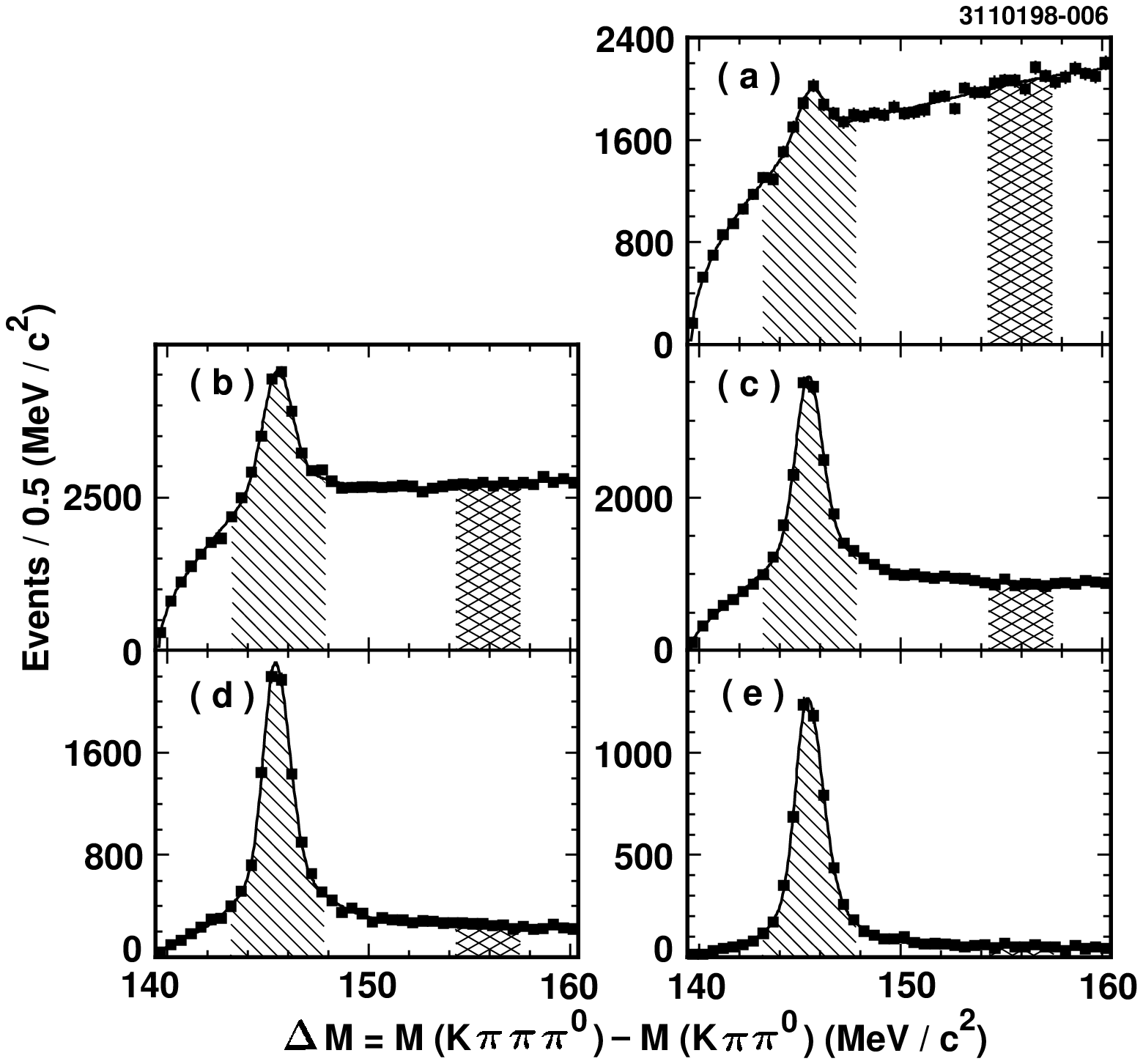,height=6.0in}}
\caption{{\label{fig:kppmsdf}}
$D^* - D$ mass difference for the $D^0 \to K \pi \pi^0$ decay mode for the 
five $x^+$ ranges  
a) $0.45<x^+<0.55$, b) $0.55<x^+<0.65$, c) $0.65<x^+<0.75$, d) $0.75<x^+<0.85$,
e) $0.85<x^+<1.0$. The solid squares are the data points and the solid
line is the fitting function as described in Section IV. The hatched area
is the signal region while the cross-hatched region is the sideband.}
\end{figure}

For each $x^+$ range, a sideband subtraction is performed. The sideband 
region 
is from 9 MeV/$c^2$ to 12 MeV/$c^2$ above the mean of the $\Delta M$ 
peak and the ratio for the sideband subtraction is determined 
by fitting the data with a bifurcated double Gaussian for the signal 
plus a background function $A+B(\Delta M)^{1/2}+C(\Delta M)^{3/2}$ 
and integrating the background shape
for the signal and sideband regions. The fits used to determine the sideband 
ratios are shown in Figures \ref{fig:kpmsdf} and \ref{fig:kppmsdf}.

The sideband-subtracted $M(K\pi)$ data is fit for each $x^+$ bin with a 
double Gaussian for the signal region plus a 
first-order polynomial background. \footnote[1]{The highest $x^+$ bin is fit
with a second order polynomial for the background since the background
is not well represented by a straight line.} 
Each of these $x^+$ bins is broken up 
into five equal $\cos \theta$ bins, where $\theta$ is the angle defined in
Section II. To prevent the fitted signal shape from having large 
fluctuations due to lower statistics in some $\cos \theta$ bins, the $D^0$ 
mass peak is fit for each $\cos \theta$ bin with the ratios of areas and
widths of the double Gaussian fixed to those found when fitting the mass 
peak in that momentum range for the entire $\cos \theta$ spectrum.

\section{Efficiencies}
It is important to understand the relative efficiencies of detecting a 
$D^{*+}$ in the various $\cos \theta$ bins. In the lowest momentum bins,
for example, the efficiency decreases as $\cos \theta$ approaches one because 
of the increased difficulty in measuring the track of a slow pion that is 
emitted in the direction opposite the $D^*$ direction in the lab
frame. Detection efficiency as a function of $x^+$ and $\cos \theta$ was 
measured by analyzing Monte Carlo data with a GEANT-based detector simulation. 

Monte Carlo events were generated using the Lund Jetset 7.3 program, where 
the $e^+ e^-$ annihilation was required to result in a $c \bar{c}$ pair with 
one of the charm quarks hadronizing to a $D^{*+}$ that decays to $D^0\pi^+$ 
with $D^0\to K^-\pi^+(\pi^0)$, while no constraints were placed on the other 
charm quark.
The $D^*$ mesons were produced such that their decay to $D^0 \pi^+$ had 
an isotropic angular distribution in the rest frame of the $D^{*+}$.

\section{Results}

\begin{figure}
\centerline{\epsfig{file=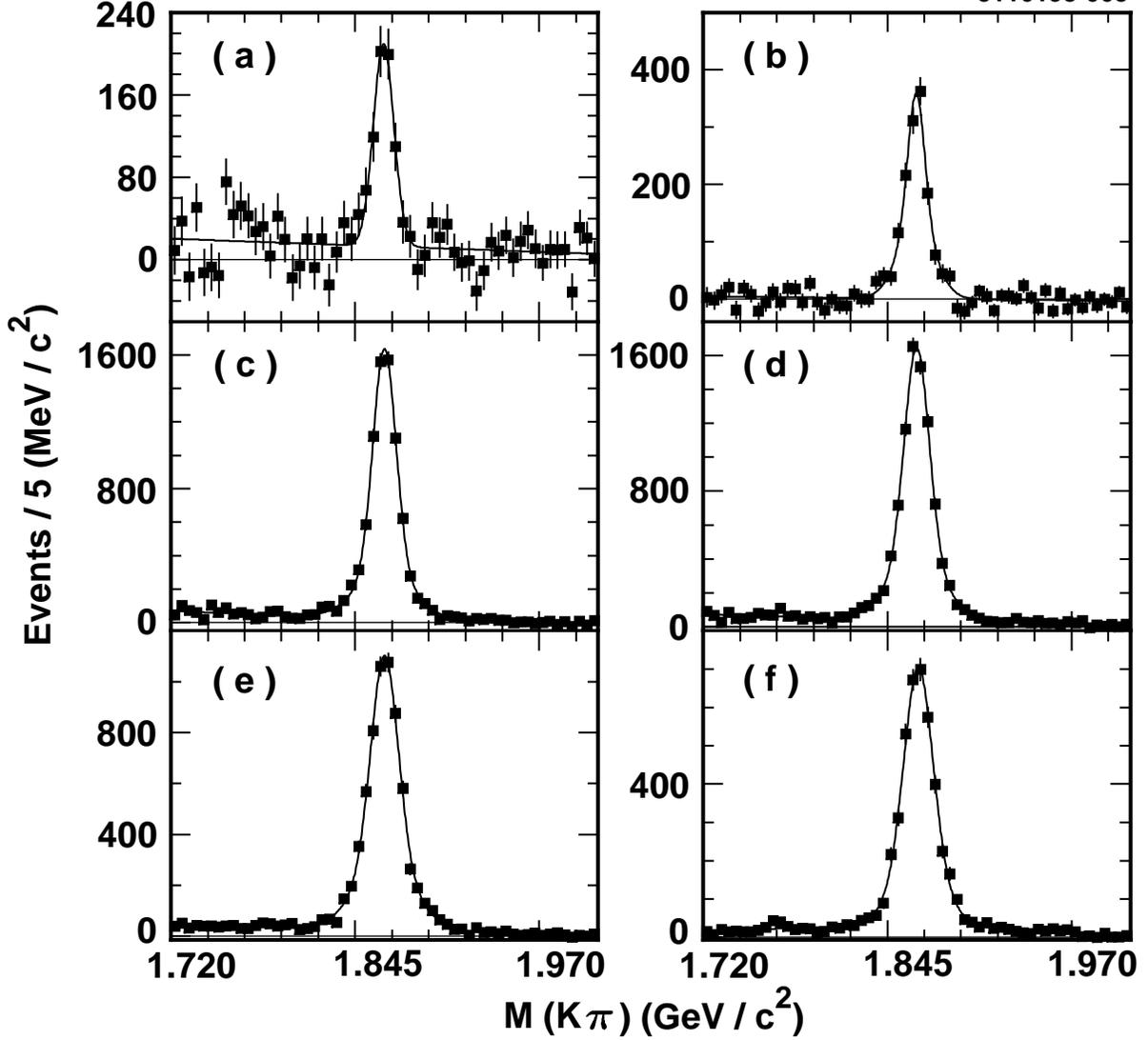,height=6.0in}}
\caption{{\label{fig:kp}}
$M(K \pi)$ after sideband subtraction for the six $x^+$ ranges 
a) $0.25<x^+<0.45$, 
b) $0.45<x^+<0.55$, c) $0.55<x^+<0.65$, d) $0.65<x^+<0.75$, e) $0.75<x^+<0.85$,
f) $0.85<x^+<1.0$. The solid squares are the data points and the solid
line is the fitting function as described in Section IV.}
\end{figure}

\begin{figure}
\centerline{\epsfig{file=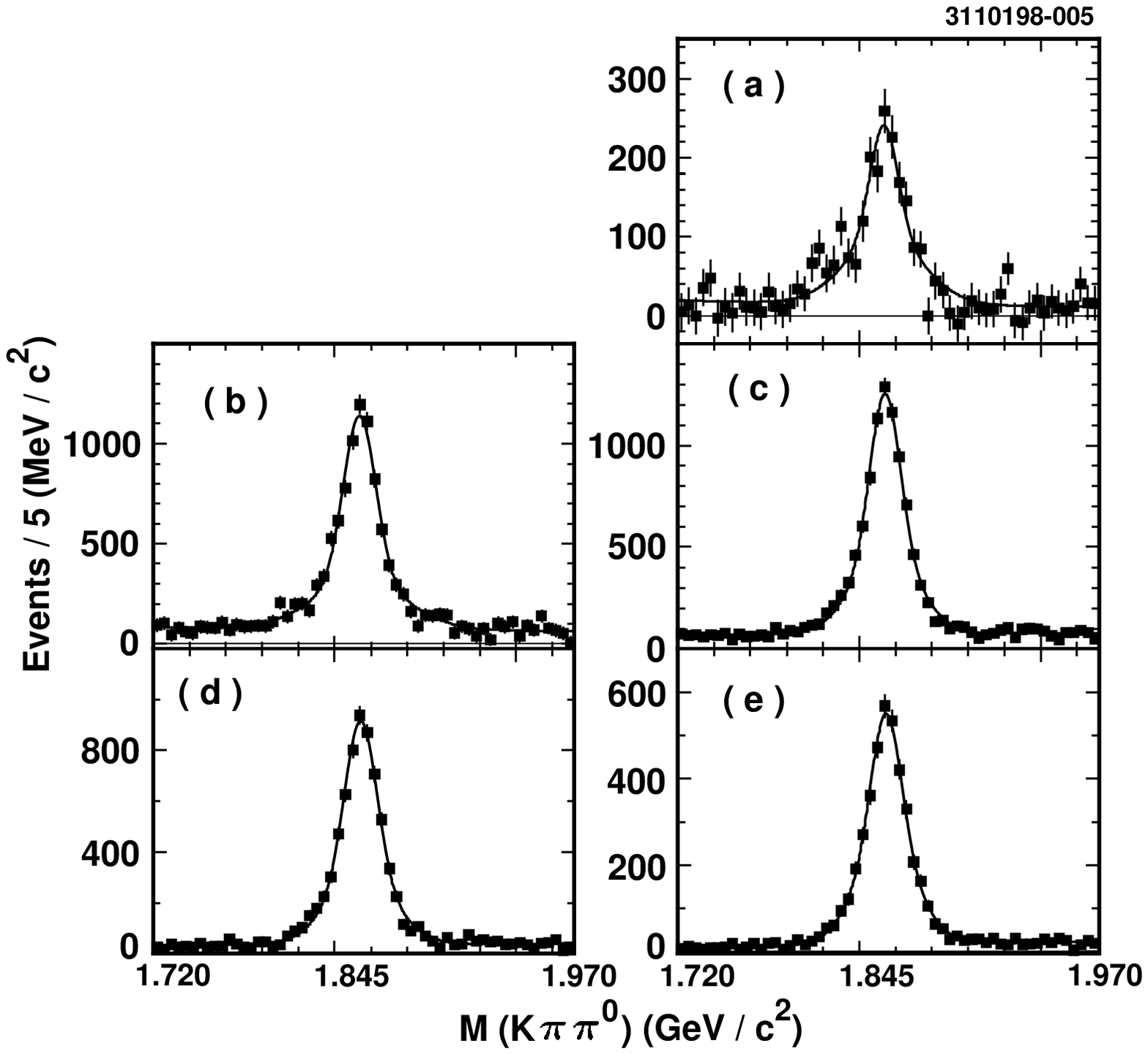,height=6.0in}}
\caption{{\label{fig:kpp}}
$M(K \pi \pi^0)$ after sideband subtraction for the five $x^+$ ranges 
a) $0.45<x^+<0.55$, b) $0.55<x^+<0.65$, c) $0.65<x^+<0.75$, d) $0.75<x^+<0.85$,
e) $0.85<x^+<1.0$. The solid squares are the data points and the solid
line is the fitting function as described in Section IV.}
\end{figure}

\begin{figure}
\centerline{\epsfig{file=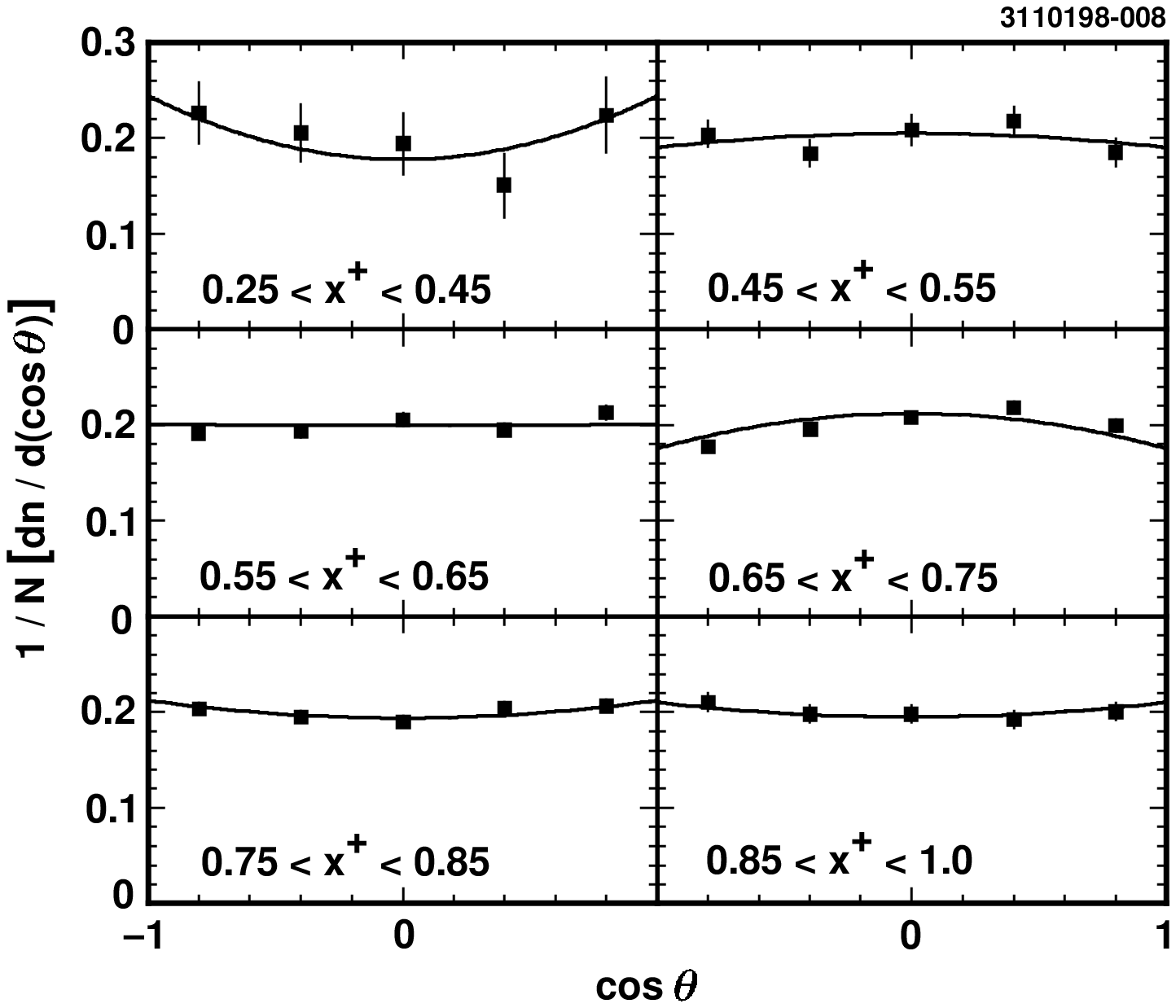,height=6.0in}}
\caption{{\label{fig:xbins}}
Normalized $\cos \theta$ distributions in the six $x^+$ ranges for the 
$D^0 \to K^- \pi^+$ and $D^0 \to K^- \pi^+ \pi^0$ decay
modes combined. The solid squares are the efficiency-corrected
yields for each $\cos \theta$ bin in the specified $x^+$ range.
These distributions are fit with the function
$W(\cos \theta)=0.4N(1+ \alpha \cos^2 \theta)$, where the factor of 0.4 
is the bin width and $N=3/(6+2\alpha)$.}
\end{figure}

The fits of the sideband subtracted $M(K\pi)$ and $M(K \pi \pi^0 )$ 
distributions for all scaled
momentum ranges are shown in Figures \ref{fig:kp} and \ref{fig:kpp}.\footnote[2]{Only the highest five momentum bins were used for the $D^0 \to
K \pi \pi^0$ mode due to the small number of signal events and low 
signal-to-noise ratio in the lowest $x^+$ range.} 
The efficiency-corrected angular distributions for both decay modes
were combined in each $x^+$ bin with a weighted average and are shown 
in Figure \ref{fig:xbins}, where 
they have each been normalized to unit area and fit with Eq. (5). 

\begin{figure}
\centerline{\epsfig{file=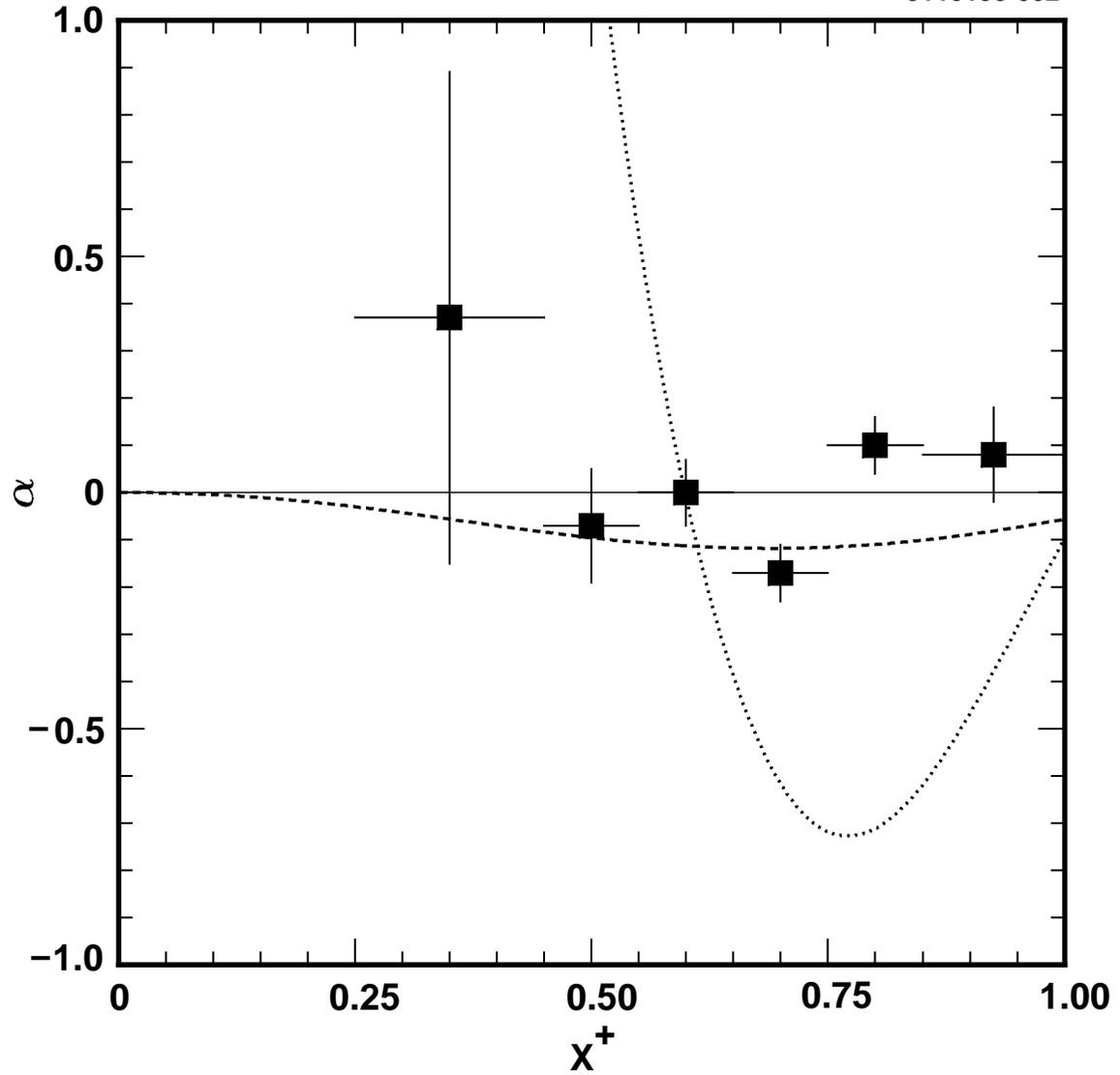,height=6.0in}}
\caption{{\label{fig:alpha}}
The values of $\alpha$ for each momentum bin are represented by the solid 
squares. Errors shown are the statistical and
systematic errors added in quadrature. The solid line represents the 
statistical model. The dotted line represents the
function predicted by Suzuki[22]. The dashed line is the 
function of Cheung and Yuan[23].}
\end{figure}

The values of $\alpha$ resulting from these fits as well as the
fits for each of the two decay modes treated separately are listed in Table
\ref{tab:alpha}. Figure \ref{fig:alpha} shows the combined results for
$\alpha$ plotted as a function of momentum as well as the theoretical
curves suggested by Suzuki \cite{oldmodel} and Cheung and Yuan \cite{newmodel}.
Table \ref{tab:rho} lists the values of $\rho_{00}$ as calculated 
from the measurement of $\alpha$ for each scaled momentum bin.
Averaging the $\cos \theta$ distributions over all momenta and then fitting
gives a value $\bar{\alpha}=-0.028 \pm 0.026$, corresponding to 
$\bar{\rho}_{00}=0.327 \pm 0.006$.

\begin{table}
\centering
\begin{tabular}{ r@{--}l r@{$\pm$}l r@{$\pm$}l r@{$\pm$}l r@{$\pm$}l r@{$\pm$}c@{$\pm$}l c }
\multicolumn{2}{c}{ } &\multicolumn{4}{c}{$D^0 \to K^- \pi^+$ }&\multicolumn{4}{c}{$D^0 \to K^- \pi^+ \pi^0$}&\multicolumn{3}{c}{Combined}&Confidence\\
\multicolumn{2}{c}{$x^+$}&\multicolumn{2}{c}{Events}&\multicolumn{2}{c}{$\alpha$}&\multicolumn{2}{c}{Events}&\multicolumn{2}{c}{$\alpha$}&\multicolumn{3}{c}{$\alpha$}&Level(\%)\\ \hline
0.25&0.45 & 687&62 & 0.37&0.35  &\multicolumn{2}{c}{ }&\multicolumn{2}{c}{ }& 0.37&0.35&0.38 &90 \\ 
0.45&0.55 & 1472&58 & -0.14&0.13 & 1830&171 & 0.09&0.24 & -0.07&0.11&0.05 & 43\\ 
0.55&0.65 & 7640&125 & 0.14&0.08 & 8305&290 & -0.18&0.08 & 0.00&0.05&0.05 & 11 \\ 
0.65&0.75 & 8432&116 & -0.13&0.06 & 8355&165 & -0.22&0.06 & -0.17&0.04&0.04 & 1 \\ 
0.75&0.85 & 6264&97 & 0.14&0.08  & 6339&118 & 0.05&0.08 & 0.10&0.05&0.02 & 73 \\ 
0.85&1.0 & 3828&83 & 0.17&0.12  & 3740&91 & -0.02&0.11 & 0.08&0.08&0.07  & 90  
\end{tabular}
\caption{{\label{tab:alpha}}
Values of $\alpha$ for different momentum ranges. First error given is
statistical, second is systematic. The last column is the confidence level of
the fit for the combined values of $\alpha$.}
\end{table}

\begin{table}
\centering
\begin{tabular}{ r@{--}l r@{$\pm$}c@{$\pm$}l}
\multicolumn{2}{c}{$x^+$}&\multicolumn{3}{c}{$\rho_{00}$}\\ \hline
0.25&0.25 & 0.40 & 0.07 & 0.07 \\
0.45&0.55 & 0.31 & 0.03 & 0.01 \\
0.55&0.65 & 0.33 & 0.01 & 0.01 \\
0.65&0.75 & 0.30 & 0.01 & 0.01 \\
0.75&0.85 & 0.35 & 0.01 & 0.01 \\
0.85&1.0  & 0.35 & 0.02 & 0.01
\end{tabular}
\caption{{\label{tab:rho}}
Values of $\rho_{00}$ for different momentum ranges. 
The first error given is statistical, the second is systematic.}
\end{table}

Similar analyses have been done by the HRS, TPC, SLD and OPAL collaborations 
\cite{matrix,align1,align5,align4}, as well as by
CLEO using a previous data set\cite{align2}.
The average values of $\alpha$ and $\rho_{00}$ in each study are presented 
in Table~\ref{tab:other}.

\begin{table}
\centering
\begin{tabular}{ l c c c}
Collaboration&$\sqrt{s}$ (GeV)&$\bar{\alpha}$&$\bar{\rho}_{00}$ \\ \hline
HRS& 29 &0.18 $\pm$ 0.08 & 0.371 $\pm$ 0.016\\
TPC& 29 &-0.14 $\pm$ 0.17 $\pm$ 0.03 & 0.301 $\pm$ 0.042 $\pm$ 0.007\\
SLD& 91 &0.019 $\pm$ 0.378 $\pm$ 0.582 & 0.34 $\pm$ 0.08 $\pm$ 0.13\\
OPAL& 91 &0.33 $\pm$ 0.11 & 0.40 $\pm$ 0.02 \\
CLEO I.5 & 10.5 &0.08 $\pm$ 0.07 $\pm$ 0.04 & 0.351 $\pm$ 0.015 $\pm$ 0.008\\
{\bf CLEO II}&{\bf 10.5}&{\bf -0.028 $\pm$ 0.026 } & {\bf 0.327 $\pm$ 0.006}\\
\end{tabular}
\caption{{\label{tab:other}}
Results for $\bar{\alpha}$ and $\bar{\rho_{00}}$ found by various 
collaborations.}
\end{table}

\section{Systematic Error}

\begin{figure}
\centerline{\epsfig{file=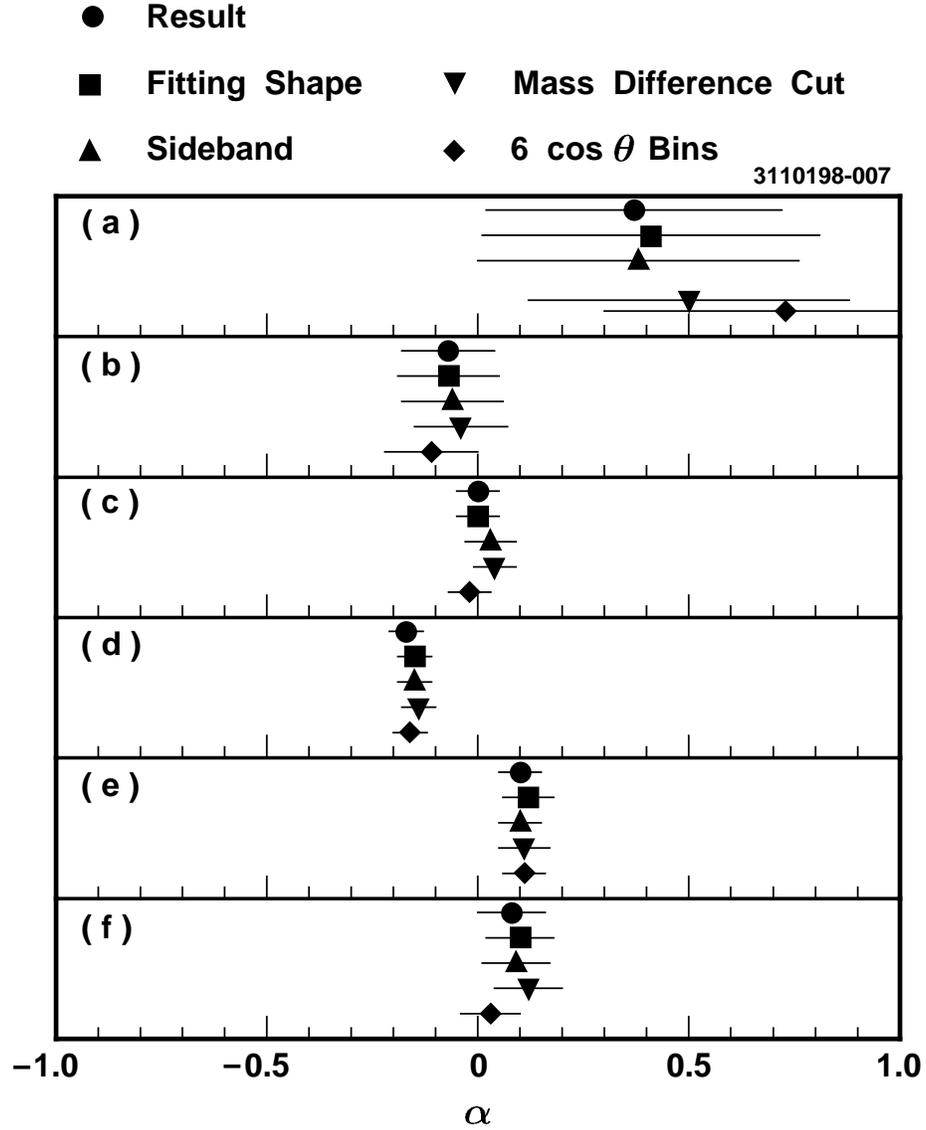,height=6.0in}}
\caption{{\label{fig:syst}}
The results from the systematic error studies for the six $x^+$ bins 
a) $0.25<x^+<0.45$, b) $0.45<x^+<0.55$, c) $0.55<x^+<0.65$, d) $0.65<x^+<0.75$, 
e) $0.75<x^+<0.85$, f) $0.85<x^+<1.0$. The different symbols represent the 
resulting values of $\alpha$ using the modifications in the analysis 
procedure as described in Section VII.}
\end{figure}

Many possible sources of absolute systematic uncertainty, such as the
overall track-finding efficiency, do not have a significant effect
on this analysis because the extraction of $\alpha$ in each momentum
range involves only the relative comparisons of the same measured 
quantity, namely the yield of the $D^0$ decays, in the different bins
of $\cos\theta$. The remaining sources of uncertainty will therefore be
related to extracting the yield and the efficiency as a function of 
$\cos\theta$. The effects of the various sources
of systematic error are shown in Figure \ref{fig:syst} while
the methods used to determine these errors are described below.

The Monte Carlo contribution to the systematic error
was accounted for by including the error in the Monte Carlo efficiencies in
the calculations of the yields. To investigate the systematic
error associated with the fitting function, the analysis was done using a
single Gaussian rather than a double Gaussian to fit the signal peaks.  
Likewise, to investigate the systematic error 
associated with the choice of range for the sideband subtraction, the analysis
was done using a sideband region from 6 MeV/$c^2$ to 9 MeV/$c^2$ above the
nominal $D^*-D$ mass difference rather than 9 MeV/$c^2$ to 12 MeV/$c^2$ 
above the nominal value. The effect
of the mass difference requirement was investigated by 
constraining the mass difference to be within 1.25 MeV/$c^2$ of
the PDG value rather than 2.5 MeV/$c^2$. The systematic effects of the 
$\cos \theta$ binning were studied by using six equal $\cos \theta$ bins 
rather than five. The differences between the 
resulting values of $\alpha$ and the central value were all summed in 
quadrature as an estimate 
of the systematic error and are included in the error bars shown in Figure
\ref{fig:alpha}. 

A small linear component in the angular distribution can easily be seen
in Figure \ref{fig:xbins} for the range $0.65<x^+ <0.75$. This is
most likely due to a slight inaccuracy in the efficiency correction from 
the Monte Carlo data. The data in Figure \ref{fig:xbins}
were fit with a straight line added to Eq. (4) as a check and
the difference in the fitted values of $\alpha$ was negligible. 

\section{Interpretation of Results}

We have measured the spin alignment of all
$D^*$ mesons produced in $e^+e^- \to q \bar{q}$ interactions at 
$\sqrt{s}=10.5$ GeV.
Although the details of the analysis ensure that the measured $D^*$ does
not come from a decaying $B$ meson, we cannot determine any other details
about the production hierarchy. From a theoretical standpoint, we are
particularly interested in the $D^*$ mesons that are produced directly
in the $e^+e^-$ collision, but we cannot distinguish these from secondary
$D^*$'s resulting from decays of charm mesons with $L>0$.
\cite{dds1,dds2,dds3}. 

The most prominent excited charm mesons, which are commonly referred to 
as $D^{**}$ mesons, consist of a charm quark and a light anti-quark
with relative orbital angular 
momentum $L=1$. They are categorized into four states with spin-parity
$J^P$ = $0^+$, $1^+$, $1^+$, and $2^+$. A $0^+$ state decay to $D^* \pi$ is 
forbidden due to spin-parity conservation while other $D^*$ modes are
expected to be suppressed. When a $2^+$ state decays 
through a $D^*$ channel, it can only produce a $D^*$ meson with a 
helicity of $\pm 1$ in 
the $2^+$ rest frame, while the $1^+$ states only decay through $D^*$ channels 
and favor a helicity of 0 in the $1^+$ rest frame. From the measurements 
available \cite{dds4,dds5}, we estimate that 16-20\% of $D^*$ mesons 
observed at CLEO could be daughters of a $D^{**}$ meson, not including the 
contribution from $D_s^{**}$ mesons.

Although the favored helicities of $D^*$'s from the decays of $2^+$ and $1^+$
charm states partially cancel, it is probable that these $D^*$'s are aligned
in their production rest frame, i.e. the rest frame of the parent $D^{**}$. 
It is expected that any effect would be most noticeable for the highest $x^+$ 
bins which has the largest correlation between the $D^*$ 4-momentum in the 
lab frame and the $D^*$ 4-momentum in the $D^{**}$ rest frame.
If the 4-momenta in the two reference frames
are uncorrelated, as tends to be the case for the lower $x^+$ bins, any
alignment of $D^*$'s from $D^{**}$'s would not be noticeable in the lab 
frame.

Due to the current lack of information about the production and decay of
$P$-wave charm meson states, we can only state that $D^{**}$ 
decays could have a significant effect on this $D^*$ spin alignment 
measurement in at least some of the $x^+$ bins.

\section{Conclusion}
This analysis is the most precise measurement of the 
spin alignment of $D^{*+}$ mesons to date. The data, without
any corrections for $D^{**}$ effects on the measurements,
agree well with the statistical model expectation that
the $J_z=0$ state has a $\frac{1}{3}$ probability of being populated.

\section{Acknowledgments}
We gratefully acknowledge the effort of the CESR staff in providing us with
excellent luminosity and running conditions.
J.P.A., J.R.P., and I.P.J.S. thank                                           
the NYI program of the NSF, 
M.S. thanks the PFF program of the NSF,
K.K.G., M.S., H.N.N., T.S., and H.Y. thank the
OJI program of DOE, 
J.R.P., K.H., M.S. and V.S. thank the A.P. Sloan Foundation,
M.S. thanks Research Corporation, 
and S.D. thanks the Swiss National Science Foundation 
for support.
This work was supported by the National Science Foundation, the
U.S. Department of Energy, and the Natural Sciences and Engineering Research 
Council of Canada.

\end{document}